\documentstyle[12pt]{article}
\title{Mathematical Techniques for Quantum Communication Theory${}^\ast$}
\author{Christopher A.~Fuchs and Carlton M.~Caves\\
\small\it Center for Advanced Studies, Department of Physics and Astronomy,\\ 
\small\it University of New Mexico, Albuquerque, NM 87131--1156}
\date{}

\begin{document}
\maketitle

\begin{abstract}
We present mathematical techniques for addressing two closely related questions
in quantum communication theory.  In particular, we give a statistically
motivated derivation of the Bures-Uhlmann measure of distinguishability for
density operators, and we present a simplified proof of the Holevo upper bound
to the mutual information of quantum communication channels.  Both derivations
give rise to novel quantum measurements.
\end{abstract}

\def\openone{\leavevmode\hbox{\small1\kern-3.8pt\normalsize1}}%

\section{Introduction}
Suppose a quantum system is secretly prepared in one of two known, but
non-orthogonal---or even mixed---quantum states $\hat\rho_0$ and
$\hat\rho_1$.
Because of the fundamental indeterminism of quantum mechanics, there is no way
to discern reliably via measurement which of the two states
has actually been prepared.
One can still ask, however, which measurement among all possible quantum
measurements will have an outcome that most likely distinguishes the one
preparation from the other?  Or, which measurement will
gather the most Shannon information about the preparation if prior probabilities
for the preparations are at hand?  
These questions, though not identical, are typical of quantum communication
theory and contain to some extent the same mathematical difficulties.
Here we develop mathematical techniques for addressing
both questions.

In Section II
we tackle a particular version of the first question by giving
a statistically motivated derivation of the Bures-Uhlmann
\cite{Bures,Uhlmann,Jozsa}
measure of distinguishability for density operators
and exploring the new quantum measurement that thus appears.
In Section III we make progress toward the second question by simplifying the
derivation of the Holevo upper bound
\cite{Holevo,Yuen,Hall,Levitin}
on the maximum mutual information for binary quantum communication 
channels; by way of this, we find a measurement that often comes close to 
attaining the actual maximum value.  

\section{Statistical Distinguishability and Fidelity}

Consider two distinct probability distributions $p_{0b}$ and $p_{1b}$
$(b=1,\ldots,N)$ for an experiment with $N$ outcomes.  Two common measures of
the statistical distinguishability of these distributions are the
Kullback-Leibler divergence or relative information
\cite{Kullback,Kailath,Cover},
\begin{equation}
K(p_0/p_1)=\sum_{b=1}^N p_{0b}\ln\!\left({p_{0b}
\over p_{1b}}\right)\;,
\label{KullbackDiv}
\end{equation}
and the Bhattacharyya-Wootters distance
\cite{Kailath,Bhatta,Wootters},
\begin{equation}
B(p_0,p_1)=\cos^{-1}\!\left(\,\sum_{b=1}^N\sqrt{p_{0b}}\sqrt{p_{1b}}\right)\;.
\label{WoottersDist}
\end{equation}
Both of these quantities take on a minimum value of zero if and only if the
distributions are not distinguishable at all, i.e.,
$p_{0b}=p_{1b}$ for all $b$, but they define different notions of maximal
distinguishability. A point of similarity between these measures is that
when $p_{1b}=p_{0b}+\delta p_b$, to lowest order both are proportional to 
powers of the Fisher information
\cite{Wootters,Fisher,Rao,Kullback2}
at the point $p_0$ in the probability simplex,
\begin{equation}
ds^2=\sum_{b=1}^N{(\delta p_b)^2\over p_{0b}}\;.
\label{LineEl}
\end{equation}
This quantity places the ultimate limit on convergence in maximum likelihood
parameter estimation \cite{Cover,Cramer} and has recently found quite a use
itself within the quantum context \cite{Braun1,Braun2,Braun3}.
If the probability simplex is thought of as a Riemannian manifold with
line element given by Eq.~(\ref{LineEl}), the Bhattacharyya-Wootters distance
is just the geodesic distance between the points $p_0$ and $p_1$
\cite{Wootters,Rao}.

The problem of statistically distinguishing the states
$\hat\rho_0$ and $\hat\rho_1$ via a quantum measurement boils down to using a
measurement with $N$ outcomes (though $N$ can be arbitrary) to generate the 
probability distributions $p_0$ and $p_1$ used in the measures (\ref{KullbackDiv}) and (\ref{WoottersDist}).  The optimal quantum measurement
with respect to either the Kullback-Leibler or Bhattacharyya-Wootters 
distinguishability measure is just that measurement which makes either of the 
respective quantities as large as it can possibly be.

These ideas are made precise through a formalization of the most general
measurements allowed by quantum theory, the positive-operator-valued measures
(POVM) \cite{Kraus}.  A POVM is a set of non-negative, Hermitian operators
$\hat E_b$ which are complete in the sense that
$ 
\sum_b\hat E_b=\hat{\openone}=\bigl(\mbox{identity operator}\bigr).
$
The subscript $b$ here, as before, indexes the possible outcomes of the
measurement.  Naturally, the conditions on the $\hat E_b$ are those 
necessary and sufficient for the standard expression
${\rm tr}\bigl(\hat\rho\hat E_b\bigr)$
to be a valid probability distribution for the $b$.
The optimal distinguishability
measurements $\{\hat E_b^{\rm K}\}$ and $\{\hat E_b^{\rm B}\}$ for the states 
$\hat\rho_0$ and $\hat\rho_1$ with respect to the measures (\ref{KullbackDiv})
and (\ref{WoottersDist}) are just those which attain
\begin{equation}
K(\hat\rho_0/\hat\rho_1)\equiv
\max_{\{\hat E_b\}}\,
\sum_b \Bigr({\rm tr}\hat\rho_0\hat E_b\Bigl)\,
\ln\!\left({
{\rm tr}\hat\rho_0\hat E_b
\over
{\rm tr}\hat\rho_1\hat E_b
}\right)\;,
\label{QKullback}
\end{equation}
and
\begin{equation}
B(\hat\rho_0,\hat\rho_1)\equiv
\max_{\{\hat E_b\}}\:
\cos^{-1}\!\left(\,\sum_b
\sqrt{{\rm tr}\hat\rho_0\hat E_b}
\sqrt{{\rm tr}\hat\rho_1\hat E_b}\right)\;.
\label{QWootters}
\end{equation}
Notice again that the number $N$ of measurement outcomes in these definitions
has not been fixed at the outset as it is in the classical
expressions~(\ref{KullbackDiv}) and (\ref{WoottersDist}).

The difficulty that crops up in extremizing quantities like
Eqs.~(\ref{QKullback}) and (\ref{QWootters}) is that, so far at least, there
seems to be no way to make the problem amenable to a variational approach: the
problems associated with allowing $N$ to be arbitrary while enforcing the
constraints on positivity and completeness for the $\hat E_b$ appear to be
intractable.  New methods are required.  Fortunately, the
Bhattacharyya-Wootters distinguishability measure
appears to be ``algebraic'' enough that one might well imagine using standard
operator inequalities, such as the Schwarz inequality for operator inner
products, to aid in finding an explicit expression for
$B(\hat\rho_0,\hat\rho_1)$.  Progress toward finding
a useful expression for $K(\hat\rho_0/\hat\rho_1)$ will, for just this
reason, be impeded by the ``transcendental'' character of the logarithm in
Eq.~(\ref{QKullback}).

At this juncture we turn our focus to optimizing the Bhattacharyya-Wootters
measure of distinguishability over all quantum measurements.  For simplicity,
here and throughout the remainder of the paper, we assume the density operators
$\hat\rho_0$ and $\hat\rho_1$ to be finite dimensional and invertible.  In this
case, we shall show that
\begin{equation}
B(\hat\rho_0,\hat\rho_1)=
\cos^{-1}\!\left(
{\rm tr}\,\sqrt{\hat\rho_1^{1/2}\hat\rho_0\hat\rho_1^{1/2}}\,
\right)\;,
\label{QWootters2}
\end{equation}
where for any positive operator $\hat A$ we mean by $\hat A^{1/2}$
$\Bigl(\mbox{or }\sqrt{\hat A}\,\Bigr)$ the unique positive operator such that
$\hat A^{1/2}\hat A^{1/2}=\hat A$.  The quantity on the right hand side of
Eq.~(\ref{QWootters2})
has appeared before in slightly different forms:  as the distance function
$
d_{\rm B}^2(\hat\rho_0,\hat\rho_1)=
2-2\cos\!\left(B(\hat\rho_0,\hat\rho_1)\right)
$
of Bures
\cite{Bures,Hubner},
the generalized transition probability for mixed states
$
{\rm prob}(\hat\rho_0\!\rightarrow\!\hat\rho_1)=
\cos^2\!\left(B(\hat\rho_0,\hat\rho_1)\right)
$
of Uhlmann \cite{Uhlmann}, and (in the same form as Uhlmann's) Jozsa's
criterion \cite{Jozsa} for fidelity of signals in a quantum communication
channel.  Moreover, in a roundabout way through the mathematical-physics 
literature (cf., for instance, in logical order
\cite{Wootters}, \cite{Hadji82}, \cite{Araki}, \cite{Hadji86}, and 
\cite{Uhlmann}) one can put together a result quite similar in spirit to
Eq.~(\ref{QWootters2})---that is, a maximization like 
(\ref{QWootters}) but, instead of over all POVMs, restricted to orthogonal
projection valued measures.  What is novel here is the 
explicit statistical interpretation, the simplicity and 
generality of the derivation, and the fact that it pinpoints the measurement
by which Eq.~(\ref{QWootters2}) is attained.
The method of choice in deriving Eq.~(\ref{QWootters2}) is an application
of the Schwarz inequality in such a way that its specific conditions for 
equality can be met by a suitable measurement.  
This is of use here because the problem of maximizing the
Bhattacharyya-Wootters distance is equivalent to simply minimizing its
cosine; that is, to prove Eq.~(\ref{QWootters2}), we need to show that
\begin{equation}
\min_{\{\hat E_b\}}\sum_b
\sqrt{{\rm tr}\hat\rho_0\hat E_b}\sqrt{{\rm tr}\hat\rho_1\hat E_b}
\,=\,
{\rm tr}\,\sqrt{\hat\rho_1^{1/2}\hat\rho_0\hat\rho_1^{1/2}}\;.
\label{WannaProve}
\end{equation}

First, however, it is instructive to consider a quick and dirty, and for this
problem inappropriate, application of the Schwarz inequality;
the difficulties encountered therein point naturally toward the correct proof.
The Schwarz inequality for the operator inner product
${\rm tr}(\hat A^\dagger\hat B)$ is given by
$|{\rm tr}(\hat A^\dagger\hat B)|^2\le{\rm tr}(\hat A^\dagger\hat A)
{\rm tr}(\hat B^\dagger\hat B)$,
where equality is achieved if and only if $\hat B = \mu\hat A$ for some constant
$\mu$.  Let $\{\hat E_b\}$ be an arbitrary POVM, 
$p_{0b}={\rm tr}(\hat\rho_0\hat E_b)$, and
$p_{1b}={\rm tr}(\hat\rho_1\hat E_b)$.  By the cyclic property of the 
trace and this inequality, we must have for any $b$,
\begin{eqnarray}
\sqrt{p_{0b}}\sqrt{p_{1b}}
&=&
\sqrt{{\rm tr}\!\left(\Bigl(\hat E_b^{1/2}\hat\rho_0^{1/2}\Bigr)^\dagger
\Bigl(\hat E_b^{1/2}\hat\rho_0^{1/2}\Bigr)\right)}\:
\sqrt{{\rm tr}\!\left(\Bigl(\hat E_b^{1/2}\hat\rho_1^{1/2}\Bigr)^\dagger
\Bigl(\hat E_b^{1/2}\hat\rho_1^{1/2}\Bigr)\right)}
\nonumber\\
&\ge&
\left|
{\rm tr}\!\left(\Bigl(\hat E_b^{1/2}\hat\rho_0^{1/2}\Bigr)^\dagger
\Bigl(\hat E_b^{1/2}\hat\rho_1^{1/2}\Bigr)\right)
\right|
=
\left|
{\rm tr}\!\left(\hat\rho_0^{1/2}\hat E_b\hat\rho_1^{1/2}\right)
\right|\;.
\label{TermIneq}
\end{eqnarray}
The condition for attaining equality here is that
\begin{equation}
\hat E_b^{1/2}\hat\rho_1^{1/2}=\mu_b\hat E_b^{1/2}\hat\rho_0^{1/2}\;.
\label{EqualCond}
\end{equation}
A subscript $b$ has been placed on the constant $\mu$ as a reminder of its
dependence on the particular $\hat E_b$ in this equation.  From inequality
(\ref{TermIneq}), it follows by the linearity of the trace and the completeness
property of POVMs that
\begin{equation}
\sum_b\sqrt{p_{0b}}\sqrt{p_{1b}}
\ge
\sum_b\left|
{\rm tr}\!\left(\hat\rho_0^{1/2}\hat E_b\hat\rho_1^{1/2}\right)
\right|
\ge
\left|\sum_b
{\rm tr}\!\left(\hat\rho_0^{1/2}\hat E_b\hat\rho_1^{1/2}\right)
\right|
=
{\rm tr}\!\left(\hat\rho_0^{1/2}\hat\rho_1^{1/2}\right)\;.
\label{SumIneq}
\end{equation}

The quantity ${\rm tr}\Bigl(\hat\rho_0^{1/2}\hat\rho_1^{1/2}\Bigr)$
is thus a lower bound to $\cos\bigl(B(p_0,p_1)\bigr)$; for it to actually
be the minimum, there must be a POVM such that, for all $b$, 
Eq.~(\ref{EqualCond}) is satisfied and 
${\rm tr}\Bigl(\hat\rho_0^{1/2}\hat E_b\hat\rho_1^{1/2}\Bigr)$ is
real and non-negative (from Eq.~(\ref{SumIneq})).  These conditions,
though, cannot be fulfilled by any POVM $\{\hat E_b\}$,
except in the case that $\hat\rho_0$ and $\hat\rho_1$ commute.  This can be 
seen as follows.
Suppose $[\hat\rho_0,\hat\rho_1]\neq 0$.
Since $\hat\rho_0$ can be inverted, condition~(\ref{EqualCond}) can be written
equivalently as
\begin{equation}
\hat E_b^{1/2}\!\left(\mu_b\hat{\openone}-\hat\rho_1^{1/2}\hat\rho_0^{-1/2}\right)
=0\;.
\label{WrongEq}
\end{equation}
The only way this can be satisfied is if we take the $\hat E_b$ to
be proportional to the projectors formed from the {\it left}-eigenvectors
of $\hat\rho_1^{1/2}\hat\rho_0^{-1/2}$ and let the $\mu_b$ be the corresponding 
eigenvalues.  This is seen easily.  The operator
$\hat\rho_1^{1/2}\hat\rho_0^{-1/2}$ is a non-Hermitian operator on an
$n$-dimensional Hilbert space, say, and thus has $n$ linearly independent 
but non-orthogonal left-eigenvectors $\langle\psi_r|$ with eigenvalues
$\sigma_r$ and $n$ linearly independent but non-orthogonal 
right-eigenvectors $|\phi_q\rangle$ with eigenvalues $\lambda_q$.  
Consider the operation of 
$
\hat E_b^{1/2}\!\left(\mu_b\hat{\openone}-\hat\rho_1^{1/2}
\hat\rho_0^{-1/2}\right)
$
on $|\phi_q\rangle$.  Equation~(\ref{WrongEq}) implies
$(\mu_b-\lambda_q)\hat E_b^{1/2}|\phi_q\rangle=0$ for all $q$ and $b$. 
Assume now, for simplicity, that all the $\lambda_q$ are distinct.  If 
$\hat E_b$ is not to be identically zero, then we must have that (modulo
relabeling) $\hat E_b^{1/2}|\phi_q\rangle=0$ for all $q\neq b$ and 
$\mu_b=\lambda_q$ for $q=b$.  This means that $\hat E_b^{1/2}$ is 
proportional to the projector onto the one-dimensional subspace that 
is orthogonal to all the $|\phi_q\rangle$ for $q\ne b$.  But since
$0=\langle\psi_r|\hat\rho_1^{1/2}\hat\rho_0^{-1/2}|\phi_q\rangle
-\langle\psi_r|\hat\rho_1^{1/2}\hat\rho_0^{-1/2}|\phi_q\rangle
=(\sigma_r-\lambda_q)\langle\psi_r|\phi_q\rangle$, we have that 
(again modulo relabeling) $|\psi_r\rangle$ is orthogonal to 
$|\phi_q\rangle$ for $q\ne r$ and $\sigma_r=\lambda_q$ for $q=r$, 
and therefore $\hat E_b^{1/2}\propto|\psi_b\rangle\langle\psi_b|$.  The 
reason Eq.~(\ref{WrongEq}) cannot be satisfied by {\it any\/} POVM is just that the
$|\psi_b\rangle$ are non-orthogonal.  When the 
$|\psi_b\rangle$ are non-orthogonal, there are {\it no\/} positive constants 
$\alpha_b$ $(b=1,\ldots,n)$ such that
$\sum_b\alpha_b|\psi_b\rangle\langle\psi_b|=\hat{\openone}$.

The lesson from this example is that the na\"{\i}ve Schwarz inequality is not
enough to prove Eq.~(\ref{QWootters2}); one must be careful to ``build in''
a way to attain equality by at least one POVM.  
Plainly the way to do this is to take advantage of the invariances of
the trace operation.  In particular, in the set of inequalities~(\ref{TermIneq})
we could have first written
\begin{equation}
{\rm tr}(\hat\rho_0\hat E_b)=
{\rm tr}\!\left(\hat U\hat\rho_0^{1/2}
\hat E_b\hat\rho_0^{1/2}\hat U^\dagger\right)
\end{equation}
for any unitary operator $\hat U$.  Then, in the same manner as there, it 
follows that
\begin{equation}
\sqrt{p_{0b}}\sqrt{p_{1b}}\ge
\left|
{\rm tr}\!\left(\hat U\hat\rho_0^{1/2}\hat E_b\hat\rho_1^{1/2}\right)
\right|\;,
\end{equation}
where the condition for equality is now
$
\hat E_b^{1/2}\hat\rho_1^{1/2}=\mu_b\hat E_b^{1/2}\hat\rho_0^{1/2}
\hat U^\dagger
$,
which, because $\hat\rho_1$ is invertible, is equivalent to
\begin{equation}
\hat E_b^{1/2}\!\left(\hat{\openone}-\mu_b\hat\rho_0^{1/2}
\hat U^\dagger\hat\rho_1^{-1/2}
\right)=0\;.
\label{CorrEq}
\end{equation}
Finally, in the manner of Eq.~(\ref{SumIneq}), we get
\begin{equation}
\sum_b\sqrt{p_{0b}}\sqrt{p_{1b}}\ge
\left|{\rm tr}\!\left(\hat U\hat\rho_0^{1/2}\hat\rho_1^{1/2}\right)\right|\;.
\label{FinalIneq}
\end{equation}
The condition for equality in this is to satisfy both
Eq.~(\ref{CorrEq}) and the requirement that
${\rm tr}\Bigl(\hat U\hat\rho_0^{1/2}\hat E_b\hat\rho_1^{1/2}\Bigr)$
be real and non-negative for all $b$.
Just as in the last example, though, there can be no POVM
$\{\hat E_b\}$ that satisfies condition~(\ref{CorrEq}) {\it unless\/} the 
operator
$\hat\rho_0^{1/2}\hat U^\dagger\hat\rho_1^{-1/2}$ is Hermitian (so that its 
eigenvectors 
form a complete orthonormal basis).  An easy way to find a unitary $\hat U$ that
makes a valid solution to Eq.~(\ref{CorrEq}) possible is to note a 
completely different 
point about inequality~(\ref{FinalIneq}).  The unitary operator $\hat U$ there 
is arbitrary; if there is to be a chance of attaining equality in
(\ref{FinalIneq}), $\hat U$ had better be chosen so as to maximize
 $\left|{\rm tr}\Bigl(\hat U\hat\rho_0^{1/2}\hat\rho_1^{1/2}\Bigr)\right|$.
It turns out that that particular $\hat U$ forces 
$\hat\rho_0^{1/2}\hat U^\dagger\hat\rho_1^{-1/2}$ to be Hermitian.

To demonstrate the last point, we rely on a result from the mathematical
literature \cite{Jozsa,Fan,Schatten}:  for any operator $\hat A$,
$
\max_{\hat U}\left|{\rm tr}\Bigl(\hat U\hat A\Bigr)\right|
={\rm tr}\sqrt{\hat A^\dagger\hat A}
$,
where the maximum is taken over all unitary operators $\hat U$; the particular
$\hat U$ that gives rise to the maximum is defined by
$\hat U\hat A=\sqrt{\hat A^\dagger\hat A}$.  Thus the $\hat U$ that gives rise
to the tightest inequality in Eq.~(\ref{FinalIneq}) is
\begin{equation}
\hat U_{\rm c}\equiv\sqrt{\hat\rho_1^{1/2}\hat\rho_0\hat\rho_1^{1/2}}
\hat\rho_1^{-1/2}\hat\rho_0^{-1/2}\;,
\end{equation}
so that Eq.~(\ref{FinalIneq}) now takes the form needed to prove 
Eq.~(\ref{WannaProve}):
\begin{equation}
\sum_b\sqrt{p_{0b}}\sqrt{p_{1b}}
\ge\left|{\rm tr}\!\left(\hat U_{\rm c}
\,\hat\rho_0^{1/2}\hat\rho_1^{1/2}\right)\right|
={\rm tr}\,\sqrt{\hat\rho_1^{1/2}\hat\rho_0\hat\rho_1^{1/2}}\;.
\end{equation}
Inserting this choice for $\hat U$ into Eq.~(\ref{CorrEq}) gives the condition
\begin{equation}
\hat E_b^{1/2}\!\left(\hat{\openone}-\mu_b\hat\rho_1^{-1/2}
\sqrt{\hat\rho_1^{1/2}\hat\rho_0\hat\rho_1^{1/2}}\hat\rho_1^{-1/2}
\right)=0\;.
\label{MeasCond}
\end{equation}
The operator
\begin{equation}
\hat M\equiv\hat\rho_1^{-1/2}
\sqrt{\hat\rho_1^{1/2}\hat\rho_0\hat\rho_1^{1/2}}\hat\rho_1^{-1/2}
\end{equation}
in this equation is indeed Hermitian and also non-negative (as can be seen
immediately from its symmetry).  Thus there is a POVM 
$\{\hat E_b^{\rm B}\}$ that satisfies Eq.~(\ref{CorrEq}) for each $b$: the 
$\hat E_b^{\rm B}$ are just the projectors onto a basis that diagonalizes
$\hat M$.  Here the $\mu_b$ must be taken to
be reciprocals of $\hat M$'s eigenvalues.

With the POVM $\{\hat E_b^{\rm B}\}$, the further condition that 
$
{\rm tr}\Bigl(\hat U_{\rm c}\,\hat\rho_0^{1/2}\hat E_b\hat\rho_1^{1/2}\Bigr)
$
be real and non-negative is automatically satisfied.  Since
the eigenvalues $1/\mu_b$ of $\hat M$ are all non-negative, one finds that
\begin{equation}
{\rm tr}\Bigl(\hat U_{\rm c}\,\hat\rho_0^{1/2}
\hat E_b^{\rm B}\hat\rho_1^{1/2}\Bigr)
={\rm tr}\Bigl(\hat\rho_1\hat M\hat E_b^{\rm B}\Bigr)
={1\over\mu_b}{\rm tr}\Bigl(\hat\rho_1\hat E_b^{\rm B}\Bigr)\ge0\;.
\end{equation}
This concludes the proof of Eq.~(\ref{QWootters2}): the Bhattacharyya-Wootters
distance maximized over {\it all} quantum measurements is a simple function of
Uhlmann's transition probability.

In the remainder of this section we report a few interesting points about the
measurement specified by $\hat M$ and the quantum distinguishability measure
$B(\hat\rho_0,\hat\rho_1)$.  Equation~(\ref{WoottersDist}) defining the
Bhattacharyya-Wootters distance is clearly invariant under interchanges of
the labels 0 and 1.  Therefore it must follow that
$B(\hat\rho_0,\hat\rho_1)=B(\hat\rho_1,\hat\rho_0)$.  A neat way to see this
directly is to note that the operators 
$\hat\rho_1^{1/2}\hat\rho_0\hat\rho_1^{1/2}$ and
$\hat\rho_0^{1/2}\hat\rho_1\hat\rho_0^{1/2}$ have the same eigenvalue
spectrum.  For if $|b\rangle$ and $\lambda_b$ are an eigenvector and eigenvalue
of $\hat\rho_1^{1/2}\hat\rho_0\hat\rho_1^{1/2}$, it follows that
$$
\lambda_b\Bigl(\hat\rho_0^{1/2}\hat\rho_1^{1/2}|b\rangle\Bigr)
=
\hat\rho_0^{1/2}\hat\rho_1^{1/2}\Bigl(\lambda_b|b\rangle\Bigr)
=
\hat\rho_0^{1/2}\hat\rho_1^{1/2}\Bigl(
\hat\rho_1^{1/2}\hat\rho_0\hat\rho_1^{1/2}|b\rangle\Bigr)
=
\Bigl(\hat\rho_0^{1/2}\hat\rho_1\hat\rho_0^{1/2}\Bigr)
\Bigl(\hat\rho_0^{1/2}\hat\rho_1^{1/2}|b\rangle\Bigr)\:.
$$
Hence,
$
{\rm tr}\,\sqrt{\hat\rho_1^{1/2}\hat\rho_0\hat\rho_1^{1/2}}
=
{\rm tr}\,\sqrt{\hat\rho_0^{1/2}\hat\rho_1\hat\rho_0^{1/2}}\,
$
and so $B(\hat\rho_0,\hat\rho_1)=B(\hat\rho_1,\hat\rho_0)$.
By the same token, the derivation of Eq.~(\ref{QWootters2}) itself must remain
valid if all the 0's and 1's in it are interchanged throughout.  This,
however, would give rise to a measurement specified by a basis diagonalizing
$
\hat N\equiv\hat\rho_0^{-1/2}
\sqrt{\hat\rho_0^{1/2}\hat\rho_1\hat\rho_0^{1/2}}\hat\rho_0^{-1/2}
$.
It turns out that $\hat M$ and $\hat N$ can define the same measurement because
not only do they commute, they are inverses of each other.  This can be seen as
follows.  Let $\hat A$ be any operator and $\hat V$ be a unitary operator such
that $\hat A=\hat V\sqrt{\hat A^\dagger\hat A}$ and hence
$\sqrt{\hat A^\dagger\hat A}=\hat V^\dagger\hat A$.  Then
$\hat A^\dagger=\sqrt{\hat A^\dagger\hat A}\,\hat V^\dagger$ and
$
\Bigl(\hat V\sqrt{\hat A^\dagger\hat A}\,\hat V^\dagger\Bigr)^2
=
\Bigl(\hat V\sqrt{\hat A^\dagger\hat A}\Bigr)
\Bigl(\sqrt{\hat A^\dagger\hat A}\,\hat V^\dagger\Bigr)
=
\hat A\hat A^\dagger
$,
and therefore
$\sqrt{\hat A\hat A^\dagger}=\hat V\sqrt{\hat A^\dagger\hat A}\,\hat V^\dagger
=\hat V\hat A^\dagger$.  In particular, if 
$
\sqrt{\hat\rho_1^{1/2}\hat\rho_0\hat\rho_1^{1/2}}=
\hat U_{\rm c}\,\hat\rho_0^{1/2}\hat\rho_1^{1/2}
$,
then
$
\sqrt{\hat\rho_0^{1/2}\hat\rho_1\hat\rho_0^{1/2}}=
\hat U_{\rm c}^\dagger\,\hat\rho_1^{1/2}\hat\rho_0^{1/2}
$ and hence
\begin{equation}
\hat M\hat N=\hat\rho_1^{-1/2}\Bigl(\hat U_{\rm c}\,\hat\rho_0^{1/2}
\hat\rho_1^{1/2}\Bigr)\hat\rho_1^{-1/2}\hat\rho_0^{-1/2}
\Bigl(\hat U_{\rm c}^\dagger\,\hat\rho_1^{1/2}\hat\rho_0^{1/2}
\Bigr)\hat\rho_0^{-1/2}=\hat\rho_1^{-1/2}\hat U_{\rm c}\,
\hat U_{\rm c}^\dagger\,\hat\rho_1^{1/2}=\hat{\openone}\;.
\end{equation}

Finally, we note an interesting expression for $\hat M$'s eigenvalues that
arises from the last result.  Let the eigenvalues and eigenvectors of
$\hat M$ be denoted by $m_b$ and $|b\rangle$; in this notation
$\hat E_b^{\rm B}=|b\rangle\langle b|$.  Then we can write two expressions for
$m_b$:
\begin{equation}
m_b\langle b|\hat\rho_1|b\rangle=\langle b|\hat\rho_1\hat M|b\rangle
=\langle b|\hat\rho_1^{1/2}\hat U_{\rm c}\,\hat\rho_0^{1/2}|b\rangle\;,
\label{Last1}
\end{equation}
\begin{equation}
{1\over m_b}\langle b|\hat\rho_0|b\rangle=\langle b|\hat\rho_0\hat N|b\rangle
=\langle b|\hat\rho_0^{1/2}\hat U_{\rm c}^\dagger\,\hat\rho_1^{1/2}|b\rangle
=\left(\langle b|\hat\rho_1^{1/2}\hat U_{\rm c}\,\hat\rho_0^{1/2}|b\rangle
\right)^\ast\;.
\label{Last2}
\end{equation}
Because the left hand sides of these equations are real numbers, so are the
right hand sides; in particular, combining Eqs.~(\ref{Last1}) and (\ref{Last2}),
we get
\begin{equation}
m_b=\left({\langle b|\hat\rho_0|b\rangle\over\langle b|\hat\rho_1|b\rangle}
\right)^{1/2}
=\left({{\rm tr}\hat\rho_0\hat E_b^{\rm B}\over{\rm tr}\hat\rho_1
\hat E_b^{\rm B}}\right)^{1/2}\;.
\end{equation}
Thus the optimal measurement operator $\hat M$ for the Bhattacharyya-Wootters
distance might be considered a sort of operator analog to the classical
likelihood ratio.  This fact gives rise to an interesting expression for the
Kullback-Leibler relative information~(\ref{KullbackDiv}) between $\hat\rho_0$
and $\hat\rho_1$ with respect to this measurement:
\begin{eqnarray}
K_{\rm B}(\hat\rho_0/\hat\rho_1)
&\equiv&
\sum_b \Bigr({\rm tr}\hat\rho_0\hat E_b^{\rm B}\Bigl)\,
\ln\!\left({
{\rm tr}\hat\rho_0\hat E_b^{\rm B}
\over
{\rm tr}\hat\rho_1\hat E_b^{\rm B}
}\right)
=
2\,{\rm tr}\!\left(\hat\rho_0\sum_b(\ln\,m_b)\hat E_b^{\rm B}\right)
\nonumber\\
&=&
2\,{\rm tr}\!\left(\hat\rho_0\ln\left(
\hat\rho_1^{-1/2}
\sqrt{\hat\rho_1^{1/2}\hat\rho_0\hat\rho_1^{1/2}}
\hat\rho_1^{-1/2}\right)\right)\;.
\label{KullTrace}
\end{eqnarray}
This, of course, will generally {\it not\/} be the maximum of the 
Kullback-Leibler information over all measurements, but it does provide a
lower bound for the maximum value.  Moreover, a quantity quite similar
to this arises naturally in the context of still another measure of quantum
distinguishability studied by Braunstein and Caves \cite{Braun3}.
  
\section{Accessible Information}

A binary quantum communication channel is defined by its signal states
$\{\hat\rho_0,\hat\rho_1\}$ and their prior probabilities
$\{1-t,t\}$ $(0\le t\le1)$.
The Shannon mutual information \cite{Shannon} for the channel
with respect to a measurement $\{\hat E_b \}$ is
\begin{equation}
I(t)\equiv H(p)-(1-t)H(p_0)-tH(p_1)=(1-t)K(p_0/p)+tK(p_1/p)\;,
\label{MutualInf}
\end{equation}
where
$H(p)=-\sum_b p_b\ln p_b$ is the Shannon information of the probability
distribution $p_b$, 
$p_{0b}={\rm tr}(\hat\rho_0\hat E_b)$, $p_{1b}={\rm tr}(\hat\rho_1\hat E_b)$,
and $p_{b}={\rm tr}(\hat\rho\hat E_b)$ for 
$\hat\rho=(1-t)\hat\rho_0+t\hat\rho_1=\hat\rho_0+t\hat\Delta=
\hat\rho_1-(1-t)\hat\Delta$
with $\hat\Delta=\hat\rho_1 - \hat\rho_0$.
The {\it accessible information\/} $I_{\rm acc}(t)$
is the mutual information $I(t)$ maximized over all 
measurements $\{\hat E_b \}$.

The problems associated with actually finding $I_{\rm acc}(t)$ and the
measurement that gives rise to it are every bit as difficult as those in
maximizing the Kullback-Leibler information, perhaps more so---for
here it is not only the logarithm that confounds things, but also the fact that
$\hat\rho_0$ and $\hat\rho_1$ are ``coupled'' through the mean density operator
$\hat\rho$.  There does, at least, exist a general upper bound to
$I_{\rm acc}(t)$, due to Holevo \cite{Holevo}, but that is of little use in
pinpointing the measurement that gives rise to $I_{\rm acc}(t)$.  In what
follows, we simplify the derivation of the Holevo bound via a variation of the
methods used in the last section.  This simplification has the advantage of
specifying a measurement whose use lower bounds $I_{\rm acc}(t)$.

The Holevo upper bound to $I_{\rm acc}(t)$ is
\begin{equation}
I_{\rm acc}(t)\le S(\hat\rho)-(1-t)S(\hat\rho_0)-t S(\hat\rho_1)\equiv S(t)\;,
\label{HolevoBound}
\end{equation}
where
$
S(\hat\rho)=-{\rm tr}\bigl(\hat\rho\ln\hat\rho\bigr)=-\sum_j\lambda_j
\ln\lambda_j
$
is the von Neumann
entropy of the density operator $\hat\rho$, whose eigenvalues are $\lambda_j$.
The key to deriving it is in realizing the importance of 
properties of $I(t)$ and $S(t)$ as functions of $t$ \cite{Holevo}.
Note that $I(0)=I(1)=S(0)=S(1)=0$.  Moreover, both
$I(t)$ and $S(t)$ are downwardly convex, as can be seen by
working out their second derivatives.  For
$I(t)$ a straightforward calculation gives
\begin{equation}
I''(t)=-\sum_{b} {\; \Bigl( {\rm tr} \bigl( \hat\Delta\hat E_b \bigr) 
\Bigr)^2
\over {\rm tr} \bigl( \hat\rho\hat E_b \bigr) }\;.
\label{Fisher}
\end{equation}
For $S(t)$ it is easiest to proceed by representing
$S(\hat\rho)$ as a contour integral \cite{Poincare}
\begin{equation}
S(\hat\rho)=-{1\over 2\pi i}\oint_C
(z\ln z)\;{\rm tr}\!\left(\bigl(z\hat{\openone}
-\hat\rho\bigr)^{-1}\right)dz\;,
\end{equation}
where the contour $C$ encloses all the {\it nonzero\/} eigenvalues of
$\hat\rho$; by
differentiating within the integral and using the operator
identity $(\hat A^{-1})'=-\hat A^{-1}\hat A'\hat A^{-1}$, one finds that
\begin{equation}
S''(t)=-\sum_{\{j,k|\lambda_j +\lambda_k\neq0\}}
\Phi\bigl(\lambda_j,\lambda_k\bigr)
\bigl|\Delta_{jk}\bigr|^2\;,
\label{Balian}
\end{equation}
where 
$\Phi(x,y)=(\ln x-\ln y)/(x-y)$ if $x\neq y$, $\Phi(x,x)=1/x$,
$\Delta_{jk}=\langle j|\hat\Delta|k\rangle$,
and $|j\rangle$ is the eigenvector of $\hat\rho$ with eigenvalue $\lambda_j$.
Expressions~(\ref{Fisher}) and~(\ref{Balian}) are clearly
non-positive.

The statement that $S(t)$ is an upper bound to $I(t)$ for
{\it any\/} $t$
is equivalent to the property that,
when plotted versus $t$, the curve for $S(t)$ has
a more negative curvature than the curve for $I(t)$ (regardless of which POVM
$\{\hat E_b \}$
is used in its definition), i.e.,
$S''(t) \leq I''(t)\leq 0$ for any POVM $\{\hat E_b\}$.
The meat of the derivation is in showing this inequality.
Holevo does this by demonstrating the existence of a function $L''(t)$, 
independent of $\{\hat E_b \}$, such that
$S''(t) \leq L''(t)$ and $L''(t)\leq I''(t)$.  From this it follows,
upon enforcing the boundary condition $L(0)=L(1)=0$, that
$I_{acc}(t)\leq L(t)\leq S(t)$.

It is at this point that a fairly drastic simplification can be made
to the original proof.  An easy way to get at such a function $L''(t)$ is 
simply to minimize $I''(t)$ over all POVMs
$\{\hat E_b\}$,
and thereafter to show that $S''(t) \leq L''(t)$.
This, again, is distinctly more tractable than extremizing the mutual
information $I(t)$ itself because no logarithms appear in 
$I''(t)$; there is hope for solution by means of the Schwarz
inequality.
This approach, it turns out, generates exactly the same function $L''(t)$ as used by Holevo in the original proof, 
though the two derivations appear to have little to do with each other.
The difference of importance here is that this approach pinpoints the 
measurement that actually minimizes $I''(t)$.
This measurement, though it generally does not maximize $I(t)$ itself, 
necessarily does provide a {\it lower\/} bound $M(t)$ to the accessible 
information $I_{\rm acc}(t)$ \cite{Fuchs}.

The problem of minimizing Eq.~(\ref{Fisher}) is formally identical to the
problem considered by Braunstein and Caves \cite{Braun1}: the
expression for $-I''(t)$ is just the Fisher information of Eq.~(\ref{LineEl}).
The steps are as follows.  The idea is to think of the numerator
within the sum~(\ref{Fisher}) as analogous to the left hand side of the
Schwarz inequality.  One would like to use the Schwarz inequality in such a way
that the ${\rm tr}\bigl(\hat\rho\hat E_b\bigr)$ term in the denominator
is cancelled and only an expression linear
in $\hat E_b$ is left; for then, upon summing over the index $b$,
the completeness property for POVMs
will leave the final expression independent of the given measurement.

This can be done (at the very least) by introducing a ``lowering'' 
super-operator ${\cal G}_{\hat C}$ with the property that for any 
positive operators $\hat A$, $\hat B$, $\hat C$,
\begin{equation}
{\rm tr}\bigl(\hat A\hat B\bigr)\le
\left|{\rm tr}\!\left(\hat C\hat B\,{\cal G}_{\hat C}(\hat A)\right)\right|\;.
\label{NewR}
\end{equation}
For then one can derive by simple applications of the Schwarz inequality (just
as in Eq.~(\ref{TermIneq}))
\begin{equation}
\Bigl({\rm tr}\bigl(\hat\Delta\hat E_b\bigr)\Bigr)^2\le
\left|{\rm tr}\left(\hat\rho\hat E_b {\cal G}_{\hat\rho}
(\hat\Delta)\right)\right|^2\le
\Bigl({\rm tr}\hat\rho\hat E_b\Bigr){\rm tr}\Bigl(
\hat E_b{\cal G}_{\hat\rho}(\hat\Delta)\hat\rho\,{\cal G}_{\hat\rho}
(\hat\Delta)^\dagger\Bigr)
\label{TBound1}
\end{equation}
and
\begin{equation}
\Bigl({\rm tr}\bigl(\hat\Delta\hat E_b\bigr)\Bigr)^2\le
\left|{\rm tr}\!\left(\hat\rho^{1/2}\hat E_b {\cal G}_{\hat\rho^{1/2}}
(\hat\Delta)\right)\right|^2\le
\Bigl({\rm tr}\hat\rho\hat E_b\Bigr){\rm tr}\!\left(
\hat E_b{\cal G}_{\hat\rho^{1/2}}(\hat\Delta)\,{\cal G}_{\hat\rho^{1/2}}
(\hat\Delta)^\dagger\right)\;,
\label{TBound2}
\end{equation}
where the conditions for equality in the rightmost inequalities of these are,
respectively,
\begin{equation}
\hat E_b^{1/2}{\cal G}_{\hat\rho}(\hat\Delta)\hat\rho^{1/2}=\mu_b
\hat E_b^{1/2}\hat\rho^{1/2}
\;\;\;\;\;\;\;\;\;\;\;\;\mbox{and}\;\;\;\;\;\;\;\;\;\;\;\;
\hat E_b^{1/2}{\cal G}_{\hat\rho^{1/2}}(\hat\Delta)=
\mu_b\hat E_b^{1/2}\hat\rho^{1/2}\;.
\label{FEqCond}
\end{equation}
Using inequalities~(\ref{TBound1}) and (\ref{TBound2}) in Eq.~(\ref{Fisher}) for
$I''(t)$ immediately gives the lower bounds
\begin{equation}
I''(t)\ge-{\rm tr}\Bigl(
{\cal G}_{\hat\rho}(\hat\Delta)\hat\rho\,{\cal G}_{\hat\rho}
(\hat\Delta)^\dagger\Bigr)
\;\;\;\;\;\;\;\;\mbox{and}\;\;\;\;\;\;\;\;
I''(t)\ge-{\rm tr}\!\left(
{\cal G}_{\hat\rho^{1/2}}(\hat\Delta)\,{\cal G}_{\hat\rho^{1/2}}
(\hat\Delta)^\dagger\right)\;.
\label{SumBound}
\end{equation}

The problem now, much like in the last section, is to choose a super-operator
${\cal G}_{\hat\rho}$ in such a way that equality can be attained in
Eq.~(\ref{SumBound}).  The ``lowering'' super-operator
${\cal L}_{\hat\rho}$ that
does the trick \cite{Braun1} is defined by its action on an 
operator $\hat A$ by
\begin{equation}
{1\over2}
\left(\hat\rho{\cal L}_{\hat\rho}(\hat A)+{\cal L}_{\hat\rho}(\hat A)
\hat\rho\right)=\hat A.
\label{Lowering}
\end{equation}
In a basis $|j\rangle$ that diagonalizes $\hat\rho$, 
${\cal L}_{\hat\rho}(\hat\Delta)$ becomes
\begin{equation}
{\cal L}_{\hat\rho}(\hat\Delta)\equiv
\sum_{\{j,k|\lambda_j+\lambda_k\ne0\}}\,
{2\over \lambda_j+\lambda_k}\Delta_{jk}|j\rangle\langle k|
\;,
\label{Rinverse}
\end{equation}
which depends on the fact that $\Delta_{jk}=0$ if $\lambda_j+\lambda_k=0$.
(For further discussion of why Eq.~(\ref{Rinverse}) is the appropriate extension
of ${\cal L}_{\hat\rho}(\hat A)$
to the zero-eigenvalue subspaces of $\hat\rho$, see \cite{Braun1}; note
that ${\cal L}_{\hat\rho}$ is denoted there by
${\cal R}^{-1}_{\hat\rho}$.)  This
super-operator is easily seen, using Eq.~(\ref{Lowering}),  
to satisfy the identity that for Hermitian $\hat A$ and $\hat B$,
$
{\rm tr}(\hat A\hat B)=
{\rm Re}
\bigl[
{\rm tr}
\bigl(
\hat\rho\hat A{\cal L}_{\hat\rho}(\hat B)
\bigr)
\bigr]
$
and, hence, also to satisfy Eq.~(\ref{NewR}).
The desired optimization is via the left member of Eq.~(\ref{SumBound}):
\begin{equation}
I''(t)\ge
-{\rm tr}\bigl({\cal L}_{\hat\rho}(\hat\Delta)\hat\rho{\cal L}_{\hat\rho} 
(\hat\Delta)\bigr)
=
-{\rm tr}\bigl(\hat\Delta {\cal L}_{\hat\rho}(\hat\Delta)\bigr)
=
-\sum_{\{j,k|\lambda_j+\lambda_k\ne0\}}\,
{2\over \lambda_j+\lambda_k}\;
\bigl|\Delta_{jk}\bigr|^2\;.
\label{FisherBound}
\end{equation}
The conditions for equality in Eq.~(\ref{FisherBound}) are 
$
{\rm Im}\bigl[ {\rm tr}\bigl(\hat\rho\hat E_b{\cal L}_{\hat\rho}(\hat\Delta)
\bigr)\bigr]=0
$
for all $b$
and (from Eq.~(\ref{FEqCond}))
\begin{equation}
\hat E_b^{1/2}\!\left(\mu_b\hat{\openone}-
{\cal L}_{\hat\rho}(\hat\Delta)\right)=0\;\;\;
\mbox{for all $b$}\;.
\end{equation}
Both conditions can always be
met by choosing the operators $\hat E_b^{\rm F}$ to be 
projectors onto the basis that diagonalizes the Hermitian operator
${\cal L}_{\hat\rho}(\hat\Delta)$ 
and choosing the constants $\mu_b$ to be the eigenvalues of
${\cal L}_{\hat\rho}(\hat\Delta)$.  

The function $L''(t)$ can now be defined as
$L''(t)
=
-{\rm tr}\bigl(\hat\Delta {\cal L}_{\hat\rho}(\hat\Delta)\bigr)$.
This, as stated above, is exactly the function $L''(t)$ used by Holevo, but
obtained there by other means.  The remainder of the derivation
of Eq.~(\ref{HolevoBound}),
to show that
$S''(t)\leq L''(t)$, consists of demonstrating the arithmetic inequality
$\Phi(x,y)\ge2/(x+y)$ (see \cite{Holevo}).

Finally we focus on deriving an explicit expression for the lower bound $M(t)$.
In the manner of Eq.~(\ref{KullTrace}) the mutual information can be written as
\begin{equation}
I(t)=
{\rm tr}\biggl( (1-t)\hat\rho_0\sum_b\bigl(\ln\alpha_b\bigr)\hat E_b
+\;
t\,\hat\rho_1\sum_b\bigl(\ln\beta_b\bigr)\hat E_b\biggr)\,,
\end{equation}
where
$\alpha_b={\rm tr}\bigl(\hat\rho_0\hat E_b\bigr)/
{\rm tr}\bigl(\hat\rho\hat E_b\bigr)$ and 
$\beta_b={\rm tr}\bigl(\hat\rho_1\hat E_b\bigr)/
{\rm tr}\bigl(\hat\rho\hat E_b\bigr)$.
The lower bound $M(t)$ is defined by inserting the 
projectors $\hat E_b^{\rm F}$ onto a basis that diagonalizes 
${\cal L}_{\hat\rho}(\hat\Delta)$ into this formula.  Now a curious fact can be
used: even though 
$\hat\rho_0$ and $\hat\rho_1$
need not commute,
${\cal L}_{\hat\rho}(\hat\Delta)$, ${\cal L}_{\hat\rho}(\hat\rho_0)$, and
${\cal L}_{\hat\rho}(\hat\rho_1)$
do all commute.  This follows from the linearity of the
${\cal L}_{\hat\rho}$
super-operator:
${\cal L}_{\hat\rho}(\hat\rho_0)={\cal L}_{\hat\rho}(\hat\rho-t\hat\Delta)=
\hat{\openone}-t{\cal L}_{\hat\rho}(\hat\Delta)$
and
${\cal L}_{\hat\rho}(\hat\rho_1)={\cal L}_{\hat\rho}(\hat\rho+(1-t)\hat\Delta)=
\hat{\openone}+(1-t){\cal L}_{\hat\rho}(\hat\Delta)$.  
Thus the same projectors $\hat E_b^{\rm F}$ that diagonalize
${\cal L}_{\hat\rho}(\hat\Delta)$
also diagonalize
${\cal L}_{\hat\rho}(\hat\rho_0)$ and ${\cal L}_{\hat\rho}(\hat\rho_1)$.
With this, it immediately follows from Eq.~(\ref{Lowering})
that $\alpha_b$ and $\beta_b$ are the 
respective eigenvalues of
${\cal L}_{\hat\rho}(\hat\rho_0)$ and
${\cal L}_{\hat\rho}(\hat\rho_1)$
corresponding to the projector $\hat E_b^{\rm F}$.
Hence $M(t)$ takes the form
\begin{equation}
M(t)={\rm tr}\biggl((1-t)\,\hat\rho_0\ln
\Bigl({\cal L}_{\hat\rho}(\hat\rho_0)\Bigr)+
t\,\hat\rho_1\ln
\Bigl({\cal L}_{\hat\rho}(\hat\rho_1)\Bigr)\biggr)\,.
\end{equation}
Similarly, one can obtain another lower bound (distinct from 
Eq.~(\ref{KullTrace}))
to the maximum Kullback-Leibler
information by using the measurment basis that diagonalizes
${\cal L}_{\hat\rho_1}(\hat\rho_0)$.



\end{document}